\documentclass[12pt]{article}
\usepackage[UKenglish]{babel}
\usepackage{amssymb,amsmath}
\usepackage{lscape}
\usepackage{natbib} 
\usepackage{graphicx}
\usepackage{caption}

\def\bx{\mbox{\boldmath{$x$}}}
\def\bw{\mbox{\boldmath{$w$}}}
\def\bm{\mbox{\boldmath{$m$}}}
\def\bn{\mbox{\boldmath{$n$}}}
\def \ds{\displaystyle}
\def\wh{\widehat{w}}
\def\seb{$\widehat{\mathrm{se}}_{\mathrm {boot}}$}
\def\bc{\mbox{\boldmath$c$}}
\def\bwh{\widehat{\bw}}

\begin{document}

\parindent 0mm

\renewcommand\refname{{\normalsize{References}}} 

\begin{center}
{\bf Who's Afraid of the Wallenius Distribution?}\\
Linda M. Haines\\
{\small Department of Statistical Sciences,\\
University of Cape Town, Rondebosch 7700, South Africa.\\
email: linda.haines@uct.ac.za ~~~ Orchid: 0000-0002-8843-5353}
\end{center}

\abstract{This paper is about the use of the Wallenius noncentral hypergeometric distribution for analysing  tables generated from a sequential biased sampling scheme without replacement and thus with row margins and  sample size fixed. The parameters of the distribution are taken to be weights which are positive and sum to one and  are thus defined on a regular simplex. The approach to analysis is presented for likelihood-based  and Bayesian inference  and is illustrated by  example,  with datasets taken from the literature or generated from semi-synthetic data. 
The analysis of  the $2 \times 2$  tables using the univariate Wallenius distribution is shown to be  straightforward, with the  parameter a single weight which  translates immediately to the requisite odds and odds ratio. The analysis of   tables with more than two categories  based on the multivariate Wallenius distribution was however more nuanced than that of the two-category tables.  Specifically, some  numerical subtleties were required in order to implement the necessary calculations and a Markov chain Monte Carlo algorithm for the Bayesian analysis was  introduced.} 
\bigskip

\noindent{\bf Keywords:} {Bayesian analysis, Biased sampling, Likelihood,  Preference weights, Wallenius distribution.}

\section{Introduction}
Wallenius, in his PhD thesis at Stanford  in 1963, developed  a scheme for sequential biased sampling without replacement as a counterpoint to random sampling and represented the scheme as an urn model.  Specifically, the scheme involves an urn containing red balls of a specified weight and white balls of a specified but different weight, with the number of balls of each colour known. The balls are removed from the urn one at a time until the required sample size is reached. The probability of drawing a red ball at a given draw is equal to the ratio of the weight of the red balls remaining in the urn over the total weight of the balls still in the urn.   \citet{wall:63}  derived the  distribution of  the number of red balls drawn from the urn at the end of the sampling process. To do so, he represented the scheme as a $2\times 2$ table with fixed margins and introduced an intuitively appealing parameter, that of the  odds of sampling a red ball.  The distribution was later extended by \citet{chess:76} to an urn model with more than two balls and the odds parameter replaced by a vector of weights which are positive and sum to one, with one weight for each colour.  The two distributions are collectively referred to as the Wallenius noncentral hypergeometric distributions. The integrals embedded in the probability mass functions of the Wallenius distributions cannot be found explicitly and are exceedingly challenging to evaluate numerically. However, in 2008  Fog released the R package {\texttt BiasedUrn}, which includes a range of functions for both the univariate and multivariate Wallenius distributions. Calculations based on these functions are, except in extreme cases,  precise and stable \citep{fog:08a,fog:08b}. In addition, \citet{math:26} includes an inbuilt-function for the Wallenius distribution which can be embedded in probability calculations based on exact arithmetic. It should be emphasised here that the Fisher noncentral hypergeometric distribution is fundamentally very different from that of Wallenius. Specifically,  the sampling scheme associated with the Fisher distribution is one in which an unknown number of balls, weighted according to colour, is taken simultaneously from the urn and  the probability of drawing a specific ball of colour is then evaluated \citep{fog:08a}.

There has been sporadic interest in the Wallenius distribution and its underlying biased sampling scheme within the scientific literature. There is a broad area of research in which the weights are taken to be, in some sense, fixed. Thus, there are a number of papers which investigate the impact of a range of the weights on the generation of data, as for example in a  model for the breeding habits of birds by \citet{schmidt:15} and  in the context of portfolio management by \citet{port:20}. In other papers,  the value of the odds ratio or preference weight is taken to be an integral part of the specific study, as for example as a function of animal  step length  in wildlife movements  \citet{anmov:14} and as the propensity matrix in configuration models for networks \citep{config:21}.
In contrast, studies in which data are available and for which the preference weights are to be estimated would seem to be very few.  Both \citet{wall:63} and \citet{chess:78} included examples in their papers, with \citet{chess:78} drawing on a dataset from \citet{manly:74}. Over 20 years later, \citet{vac:00} reported the use of the  Wallenius distribution in assessing the effectiveness of a `leaky' vaccine on individuals who were vaccinated for measles and did or  did not develop the disease.  Again, after a gap of 13 years,  \citet{mauro:13} proposed a statistical mechanical model for the structure of glass and the model based on six weight parameters  was implemented in a later paper  by \citet{bodkermao:19}. In addition, \citet{fortuny:18}  used the binary version of the Wallenius distribution to quantify the preferences for  11,917 movies by 7,642 users and thereby to provide  training data for a naive Bayes prediction model. Most recently, \citet{graz:19}, in a landmark paper, demonstrated the use of the multivariate Wallenius distribution to estimate the  preference weights associated with two sets of tables extracted   from survey data. As an aside, the $2 \times 2$ tables of interest can also be regarded as contingency tables with two margins  fixed \citep{choi:15}. This interpretation is  built into the  R package {\texttt goseq} for analysing data in gene ontology, with the  Wallenius distribution as opposed to Fisher's exact test providing the basis for a null hypothesis \citep{young:10, gao:11}.

The aim of the present study is to demonstrate that datasets based on  the scheme of  biased sampling introduced by \citet{wall:63} can be analysed and the attendant preference weights estimated  in a straightforward manner within the framework of frequentist, Fisherian and Bayesian inference \citep{eh:16}.  The paper is structured as follows. The univariate and multivariate Wallenius distributions are introduced  in Section 2 and the analysis of $2 \times 2$  tables using the univariate Wallenius distribution is presented in Section 3. Numerical methods for facilitating the calculation of the probability mass functions and  an MCMC approach to simulating data from the posterior distribution of the parameters  in the Bayesian analysis for the multivariate case  are introduced in Section 4. Section 5 is devoted to the analysis of  data for single and multiple $3 \times 2$  tables extracted as semi-synthetic data from the complete dataset of \citet{graz:19}. Analyses based on the Wallenius distribution for two real world datasets  are then delineated in Section 6 and a discussion of the main results is given in Section 7.
The analyses presented in this paper were programmed in R \citep{rlang:26} using the packages \texttt{BiasedUrn} \citep{fog:26} and  \texttt {Ternary} \citep{smith:17} and in \citet{math:26}. The examples introduced draw heavily on the  paper of \citet{graz:19}.
\section{Preliminaries}\label{sec2}
\subsection{Univariate Setting}
Consider an urn containing  $m_1$ red balls each of weight $w_1 >0 $ and $m_2$ white balls each of  weight $w_2 > 0$.  Suppose that  $n$ balls are drawn one at a time without replacement from the urn and that  the probability of a ball being selected at a given draw is equal to the weight of the balls  of the same colour in the urn over the weight of all the balls in the urn. The sampling is therefore biased unless $w_1=w_2$.   The setting can be represented  as a $2 \times 2$ table with the  totals $m_1$ and  $m_2$ and the sample size $n$ fixed and  $x$  the realisation of the random variable $X$, the number of red balls drawn from the urn. The table is summarised  in Table\;\ref{tab:uni1} and  the probability of drawing a red ball can be expressed as $\ds \frac{(m_1-x_1) w_1}{(m_1-x_1) w_1+(m_2-x_2) w_2}$ and of drawing a white ball as $\ds \frac{(m_2-x_2) w_2}{(m_1-x_1) w_1+(m_2-x_2) w_2}$. \citet{wall:63} used this setting as the basis for deriving the probability mass function (pmf) of $X$ in terms of the odds of a red ball being drawn.  The odds parameter  $\beta= \frac{w_1}{w_2}$ embedded in the  expression so obtained is  invariant to scale and in the present study the pmf  is taken to be 
 \begin{equation}
\mbox{Prob}(X=x | w)=\binom{m_1}{x} \binom{m_2}{n-x} \int_0^1 (1-t^{\frac{w}{d}})^{x} (1-t^{\frac{1-w}{d}})^{{n-x}}\; dt 
    \label{eq:unipmf}
  \end{equation}
where $w$ represents the weight of a red ball and $1-w$ that of a white ball and $d=w (m_1-x)+(1-w)  \big( m_2-(n-x) \big)$.  The constraints of the hypergeometric distribution   $\max(0,n-m_1) \le x \le \min(n,m_1)$ necessarily hold. The realization $x$ is said to follow a univariate Wallenius noncentral hypergeometric distribution, denoted here as  $x \sim {\mbox W}(m_1,m_2,n,w)$. 
Suppose now that the biased sampling scheme has been implemented and the data collected.  The weight $w$ can then be interpreted as an unknown parameter with attendant likelihood $L(w | x; m_1,m_2,n)$ and is to be estimated. The weight $w$ is chosen here to be the unknown parameter rather than the odds parameter $\beta$ since functions of $w$  have a compact representation on the interval $[0,1]$ and, in addition, the weight  itself is straightforward to interpret.  The odds of drawing a red ball and the odds ratio can  immediately be recovered from the weight $w$. 
\subsection{Multivariate Setting}
Consider now the multivariate counterpart of the biased sampling scheme introduced in the previous subsection, with balls of more than two colours  in the urn.  Specifically, suppose that there are $m_i$ balls in the urn of colour $i$ and weight $w_i$, $i=1, \ldots, c$,  where $c$ denotes the number of colours and the total number of balls is given by $M=\sum_{i=1}^c m_i$.  Suppose further that a fixed number of  balls, $n < M$,  are drawn  sequentially and without replacement from the urn according to the Wallenius biased sampling scheme and  that there are $x_i$ balls of colour $i$, $ i=1, \ldots, c$,  in the resultant sample. The  scheme can be summarised compactly as the  $c \times 2$  table displayed  in Table\;\ref{tab:multi},
with the totals $m_1, \ldots, m_c$ and the sample size $n$  fixed and  the counts $x_i, i=1,\ldots, c-1,$  realisations of the random variables $X_1, \ldots ,X_{c-1}$, the numbers of balls of the  $c-1$ colours drawn in the sample from the urn. The count  $X_c=x_{c}$ is determined by $x_c = n-\sum_{i=1}^{c-1}x_i$.  \citet{chess:76}  developed an expression for the pmf of the random variables $X_1, \ldots ,X_{c-1}$   by invoking that for the univariate pmf derived by Wallenius, that is expression (\ref{eq:unipmf}), and introducing a vector of positive weights $\bw=(w_1, \ldots, w_c)$ placed on the colours.  The probability of drawing a ball of colour $i$ therefore given by  $\ds \frac{(m_i-x_1) w_i}{ \sum_{j=1}^c (m_j-x_j) w_j}, i=1,\ldots,c$ and the pmf is given by 
\begin{equation*}
\mbox{Prob}(X_{1:c}=\bx|\bm,n,\bw)=\prod_{i=1}^c \binom{m_i}{x_i} \int_0^1 \prod_{i=1}^c (1-t^{\frac{w_i}{d}})^{x_i} \; dt 
  \end{equation*}
where $X_{1:c}$ denotes the number of balls of each colour drawn, $\bm=(m_1, \ldots, m_c)$, $\bx=(x_1, \ldots, x_c)$, $\bw=(w_1, \ldots, w_c)$ and  $d=\sum_{i=1}^c w_i (m_i-x_i)$.  The realizations $\bx$  are then said to follow  the multivariate Wallenius noncentral hypergeometric distribution, that is, $\bx \sim W(\bm,n,\bw)$. \citet{chess:78}, as a counterpoint to the biased sampling scheme defining Table\;\ref{tab:uni1}, observed that  if data comprising a complete $c \times 2$ table are available,  the  weights can be taken as  unknown parameters with likelihood $L( \bw| \bx; \bm, n)$. 
Furthermore, the weights can be scaled without loss of generality and  can be taken  to sum to one, that is $0 \le w_i \le 1$ and $\sum_{i=1}^c w_i =1$. This latter setting is of interest in the present study.  
\subsection{A Multivariate Dataset}\label{s23}
\citet{graz:19} introduced an approach to analysing preference data which is based on the multivariate Wallenius noncentral hypergeometric distribution. The authors invited $174$ statisticians to choose at least 10 but no more than 20 journals which they preferred, according to specified criteria,  from a  list of $124$. After the responses were submitted, the researchers divided the journals into five categories, namely Methodology, Probability, Applied Statistics, Computation and Economics, and  counted the numbers of journals in the categories selected by each respondent.   
The data can be assembled in 174 five-by-two  tables,  with rows corresponding to the categories of  journal and  responses the selection and non-selection of a journal.  For clarity, the $5 \times 2$ table for the 20th respondent is shown in Table\;\ref{tab:gdata}(a).
The dataset provides an invaluable source of data, both real and semi-synthetic, for  the examples which now follow. Note that the term category for the rows of each table, rather than group,  is used here  to comply with the terminology of \citet{graz:19}, that is of biased sampling. 
\section{The Univariate Wallenius  Distribution}
\subsection{The Wallenius Data}
The  genetic study  reported by Wallenius in  his  1963 thesis  is introduced here. Specifically, 215  rabbit neonates born on a small island off the coast of San Fransisco were captured, blood typed as homozygote or heterozygote, and released. Three months later the surviving rabbits were recaptured and blood typed again. The results, according to blood type, are summarised in the  $2 \times 2$  table displayed in Table~\ref{tab:rabbits}. 
The question posed by the geneticists was as to whether or not the  survival of a neonate rabbit depends on phenotype. 
An objective interpretation of the table is that only the row totals are known at the beginning of the study and that the table should be regarded as a $2 \times 2$ contingency table with one fixed margin and analysed accordingly \citep{choi:15}. 
However, an alternative interpretation addresses the question ``Is Nature biased?''\:directly by taking the  table to be generated from a biased sampling scheme devised by Nature. The table can then be regarded as  a $2 \times 2$ table with two fixed margins and the univariate Wallenius distribution invoked to quantify the ``degree of bias'' embedded in the data \citep{wall:63}. 
This latter approach is undoubtedly contentious but follows that of Wallenius' thesis and  is adopted here to fix ideas. 
Thus, the values of  the totals $m_1$ and $m_2$ and the sample size $n$ in Table \ref{tab:uni1} are given by $95,120$ and $75$, respectively, and are taken to be fixed.  The random variable $X$ is taken to denote the number of homozygotes sampled and the realization of $X$, denoted $x$, is assumed to have an unknown  preference weighting of $w$ and to be distributed as $\mbox{W}(95,120,75,w)$. Note that the observation $x$ is an integer taken over a range from $0$ to $75$ and that the odds of a homozygote sampled is given by $\frac{w}{1-w}$ and the odds ratio by $\big(\frac{w}{1-w}\big)^2$.
\subsubsection{A Frequentist Lens}
 The probability $\mbox{Prob}(X \ge x|w)$ increases smoothly with the unknown  weight parameter $w$. As a consequence,  the confidence distribution and attendant confidence density for  $w$ can be constructed \citep{paw:01, xie:13} and are given by 
$$
C(w)=\mbox{Prob}(X \ge 41|w)= P_{w}(X \ge 41)
 \mbox{~and~}  c(w)=\frac{\partial P_{w}(X \ge 41)}{\partial w}. 
$$
Explicit but lengthy expressions for the confidence distribution $C(w)$ and the confidence density $c(w)$ in terms of the  weight $w$ can be found  using  exact arithmetic in \citet{math:26}.
The  mean, median and  mode of the confidence density are given by $0.619, 0.621$ and $0.624$, respectively, and the 0.025 and 0.975 percentiles by $0.509$ and $0.722$. 
\subsubsection{A Likelihood Lens}
\begin{figure}[htbp]
\centering
\includegraphics[width=\textwidth]{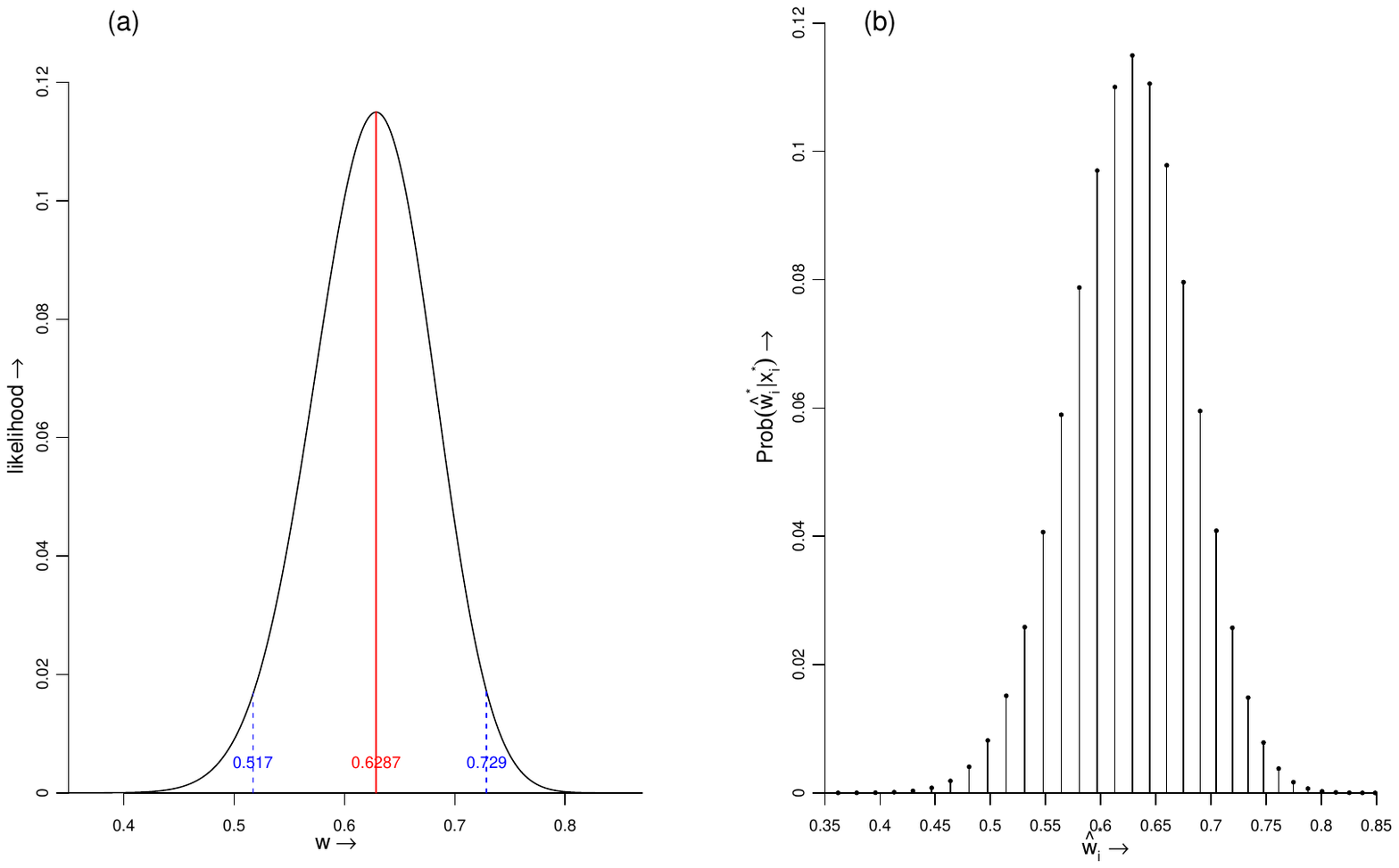} 
\caption{Wallenius data: (a) The likelihood function, the MLE in red and the limits of the 95\% Wilks' LR intervals in blue  and (b) the ideal bootstrap distribution for the replications of $\wh$.}
\label{fig:wmle}
\end{figure}

The likelihood of the weight $w$, denoted  $L(w|x)$, at the observed count $x=41$ follows immediately from the pmf of the univariate Wallenius distribution. The maximum likelihood estimate (MLE) of the parameter is given by $\hat{w}=0.6287$ and the $95\%$ confidence interval obtained by invoking Wilks' likelihood ratio by  $(0.517, 0.729)$. 
The likelihood, the MLE, $\wh$, and the $95\%$ confidence limits are shown in Figure \ref{fig:wmle}(a). 
The MLE of the odds of a  homozygote being sampled is thus given by 1.693 and that of the odds ratio by 2.867.  In addition, the pure likelihood interval for $w$ which encloses all weights for which the relative likelihood is greater than 15\%  is given by $(0.518, 0.729)$. 

A parametric bootstrap sample can be drawn from the univariate Wallenius distribution specified by W$(95,120,75,\wh)$, where $\wh=0.6287$ is the MLE of the weight $w$. A bootstrap sample is a singleton, $x^{\star}$, and the probability of drawing the singleton is given by Prob$(X=x^{\star}|\wh)$ which can be calculated exactly in \citet{math:26}. The bootstrap distribution therefore comprises values  $x_i^{\star} $ equal to the integer $i$ with associated probabilities $p_i=\mbox{Prob}(X=x_i^{\star}|\wh)$ for $i = 0, \ldots, 75$ and  is discrete and can be construed as ideal. 
The statistic of interest here is the MLE of the weight, that is $\wh$. Bootstrap estimates of the statistic can thus be obtained by invoking the likelihood $L(w | x_i^{\star})$ and finding the attendant MLE $\wh_i^{\star}$ for $ i= 0, \ldots, 75$.  As an aside, there is some ambiguity with respect to observations or bootstrap samples of $0$ and $75$. Thus, $L(w | X=0)  \uparrow 1$  as $w \to 0$ and $L(w | X=75)  \uparrow 1$  as $w \to 1$ \citep{math:26} and the   MLE for the weights can therefore be  construed as being $0$ and $1$, respectively.  On the other hand, from a practical perspective, weights $w$ of $0$ and $1$ can be taken to indicate that all the neonate rabbits captured in the survey belong to the same phenotype, an observation which has no meaning in the present context. Here, for mathematical simplicity, the estimate of $\wh_0^{\star}$ is taken to be  $0$ and that of $\wh_{75}^{\star}$ to be 1.  In fact, the attendant  bootstrap probabilities $p_0$ and $p_{75}$ are of the order of $10^{-37}$  and $10^{-29}$, respectively,  and, as a consequence,  the  samples  $x_0^{\star}$ and $x_{75}^{\star}$ have negligible impact on the  bootstrap calculations.  The  bootstrap estimate of the standard error for the MLE of the weight, that is  $\wh$, was found to be  $\widehat{\mathrm{se}}_{\infty} =  0.054$ and, following that, the standard confidence interval for $\wh$ is   given by $(0.522, 0.735)$. In addition, the percentile method for finding bootstrap confidence intervals was invoked and yielded a 95\% central percentile interval  for $w$ of $(0.514, 0.734)$.  Because of the discrete nature of the bootstrap distribution, the values of these confidence intervals did not change when the bias-correction method BC$_{a}$ was introduced. The distribution of the bootstrap replications $\wh_i^{\star}, i=0, \ldots, 75$ is shown in Figure \ref{fig:wmle}(b).
\subsubsection{A Bayesian Lens}\label{s313}
Consider now a Bayesian approach to the analysis of the univariate Wallenius data. The prior probability distribution is taken here to be the most natural candidate, that of the beta distribution, denoted Be$(a,b)$ where $a$ and $b$ are the two shape parameters. The posterior distribution of the parameter $w$  is proportional to $L(w|x) \times \pi(w)$, where $\pi(w)$ is the probability density function (pdf) of the beta distribution, and cannot be expressed in closed form. However, since the weight $w$ is the only parameter in the model, the normalising factor, which is the evidence  $\int_0^1 L(w|x) \pi(w) dw$, can be approximated to a high degree of accuracy by Monte Carlo integration and the properties of the posterior distribution thus examined \citep{speagle:19}. In the present study,  the non-informative or flat prior, Be$(1,1)$, was introduced as a benchmark and is consistent with objective Bayes inference \citep{eh:16}. In addition, the priors Be$(1,2)$ and Be$(2,4)$, which reflect a sense  that  the heterozygotes are more likely to survive than is indicated by the flat prior, were also selected and can be construed as being weakly informative and  informative, respectively. 
The requisite posterior densities were  approximated by means of a grid with a  spacing of $10^{-7}$ over the interval $(0,1)$ and the attendant  mean, standard deviation, median,  and 95\% and $68\%$ credible intervals were found. The results  are summarised in Table\;\ref{tab:unibayes}.  
The densities of the weight were very slightly skewed to the left, underscoring the nature of the attendant beta priors. The impact of the priors in terms of increasing the weight associated with the heterozygote neonates was relatively small however and  only the 95\% credible interval for the prior Be$(2,4)$ spanned a weight of $0.5$. It is worth noting that more highly informative priors, such as the beta distribution Be(2,6), do not align well with the likelihood.
\subsection{Multiple {\boldmath{$2 \times 2$} Tables}}
In order to demonstrate the use of the univariate Wallenius distribution in the analysis of multiple $2 \times 2$ sampling tables, data for two categories of journal were  extracted  as semi-synthetic data from the complete dataset of \citet{graz:19}  introduced in Section \ref{s23}.    The totals in the row margins of each individual table were taken to be those of the complete dataset but the sample sizes for each respondent, that is the total number of journals  selected in the two categories, were retained. 
 For clarity, the responses  of an individual taken from the complete dataset before and after the Methodology and Applied Statistics categories were extracted are shown, together with the respective margins,  in Table\;\ref{tab:gdata}.  The analysis of the $2 \times 2$ tables  with a single weight parameter common to all respondents is now reported for each of the ten pairs of journals.
\subsubsection{A Likelihood Lens}\label{s321}
 Consider first a likelihood-based analysis of the multiple $2 \times 2$ tables for the two categories of journal, Methodology and Applied Statistics,  with a focus on the  preference weight for Methodology,  $w$. No tables in which both preferred counts are zero were identified and the likelihood can thus be expressed as   $L_T(w|x_1, \ldots, x_T)=\prod_{h=1}^{T}L(w|x_h,n_h)$, where $T=174$ denotes the number of tables and, for the  $h$th respondent, $x_h$ denotes the number of Methodology journals preferred and $n_h$ the sample size, $h=1, \ldots, T$. The likelihood $L_T(w|x_1, \ldots, x_T)$ was found to be unimodal and the MLE of the weight, denoted $\wh$, and the attendant 95\% pure likelihood interval calculated. In addition, a nonparametric bootstrap of $10,000$ samples was conducted and the bootstrap standard error, \seb~and  the 95\% central percentile interval for the MLE  $\wh$ were obtained.  The results  are given in Table \ref{tab:pairs}.
 
It is worth pausing here, albeit briefly, to consider the calculation of the  likelihoods. The values of the individual  likelihoods $L(w|x_h,n_h), h=1,\ldots,174$, lie in the interval $[0,1]$ and the  product of the likelihoods, that is $L_T(w|x_1, \ldots, x_T)$, will therefore be extremely small. In fact, for the journal categories of interest here, $L_T(w|x_1, \ldots, x_T)$ has a maximum of the order of $10^{-163}$. Nevertheless, all calculations using \citet{math:26} and the  R package {\tt BiasedUrn} proved to be numerically stable. 

All ten pairwise comparisons of the five categories of journal, that is Methodology, Probability, Applied Statistics, Computation and Economics, were analysed following the same procedure as that detailed for Methodology and Applied Statistics  and the results are summarised in Table\:\ref{tab:pairs}.  The entries  $\wh$ and $1-\wh$  represent the MLEs of the  preference weights associated with Categories 1 and 2, respectively, and the entries for \seb~and the 95\% pure likelihood and bootstrap percentile intervals all relate to the MLE $\wh$ for Category 1.  In addition,  $T$, the total number of tables used in the analysis, and  $n_{10}$ and $n_{01}$, the numbers of times only Category 1 and only Category 2 were preferred, are also recorded in the table in order to highlight the structural differences embedded in the two-category comparisons.  It  is clear from the table that the  broad rating of the categories of journal from most to least preferred is given by Computation, Methodology,  Applied Statistics, Economics and Probability and that there are no circularities in the ordering.

The results from Table \ref{tab:pairs} can be interpreted within the context of the method of paired comparisons \citep{david:63}. Suppose that estimates of the nominal  probabilities for the five categories of journal, denoted  $\pi_i, i=1, \ldots, 5$, where $0 < \pi_i < 1$ and $\sum_{i=1}^5 \pi_i =1$, are sought. Such estimates can be obtained, albeit somewhat naively, as those values of $\pi_i$  which  minimise the sum of squares criterion $\underset{
\begin{subarray}{c}  
i,j=1, \ldots, 5 \\ 
i < j \end{subarray}}{\sum} \big( \wh_{ij}-\ds \frac{\pi_i}{\pi_i+\pi_j} \big)^2
$, where $\wh_{ij}$ is the MLE of the preference weight for category $C_i$  in a comparison  with category $C_j$ for $ i < j, i, j=1,\ldots, 5$.  The estimates of the  nominal probabilities, denoted $\hat{p}_i,i=1, \ldots, 5$,  are given by $0.301,0.039, 0.199, 0.373$ and $0.089$, in the order Methodology, Probability, Applied Statistics, Computation and Economics. Standard errors associated with the estimated probabilities were also calculated by invoking a nonparametric bootstrap procedure. Specifically,  the bootstrap was conducted by sampling the 10 pairs of  categories uniformly and exhaustively to yield all compositions of 10, that is $92,378$, bootstrap samples and the bootstrap standard errors of  the estimates $\hat{p}_i,i=1, \ldots, 5$, were found to be  $0.062, 0.023, 0.045, 0.071$ and $0.023$,  respectively. 
\subsubsection{A Bayesian Lens}\label{s322}
A Bayesian approach with a beta prior can be adopted here and follows that for the single table detailed in  Section \ref{s313}. For example, taking the prior on the weights to be Be(1,1), that is a flat prior, yielded  95\% credible intervals which are very close to the 95\% likelihood and bootstrap percentile intervals of Table\;\ref{tab:pairs} and further details are not recorded here. 
\section{Computation and an MCMC}
The parameters of the Wallenius distribution with $c$ categories are taken in this study to be weights which lie on the $c$-dimensional simplex with $c \ge 2$, denoted  $S^{c-1}$,   and are represented in barycentric coordinates as $\bw=(w_1, w_2, \ldots, w_c)$, with $0 \le w_i \le 1$ and $\sum_{i=1}^c w_i =1$ . This notion is not problematic in the analysis of data based on the univariate setting because the simplex $S^1$ is the line segment $[0,1]$ and the two weights are defined by a single parameter. However,  the use of barycentric coordinates in the  analysis of a $c \times 2$  table with more than two categories, that is  with $c >2$, which is based on the multivariate Wallenius distribution, requires further attention.  
\subsection{Calculations}
\subsubsection{Weight transformation}\label{s411}
Optimisation of functions, such as the likelihood  $L(\bw |  \bm, n, \bx )$,  with respect to the weights  $\bw$ is necessarily constrained. However, the  weights can be transformed to `working  parameters'  so that  the optimisation is then unconstrained. Here, an approach taken from that for the construction of  approximate optimal designs  is adopted, with  the transformation expressed as  
 $$
w_j=\frac{z_j^2}{\sum_{i=1}^c z_i^2}  \mbox{~and thus as ~}   z_j=\sqrt{w_j} \mbox{~~where ~} z_j \in {\mathbb{R}} \mbox{~and~}  j=1, \ldots, c,
$$  
and embedded  into routines for optimisation, such as {\tt optim} in base R  \citep[p.~130]{adt:07}.  
There are a number of other  strategies in R  for optimising a function with respect to weights which lie in the interval $[0,1]$ and sum to one but these were not explored in the present study. 
\subsubsection{Cartesian and Barycentric Coordinates}
Weights expressed in barycentric coordinates are intuitively meaningful. For example,  surfaces and regions  on the two- and three-dimensional simplexes, that is the triangle and the tetrahedron, can be represented graphically. But certain operations, such as moving a weight within a simplex, is not valid \citep{terv:13}. For example, the two-dimensional simplex in barycentric coordinates has zero volume in three-dimensions.  In such cases however, it is possible to move seamlessly between barycentric and  Cartesian coordinates and, thereby, to avoid  any such complications. This approach can be  implemented by invoking  functions available in the R package {\tt geometry}.  As an aside, it is worth noting that the geometry of the two-dimensional simplex is not widely documented \citep{sc:12} and that for simplexes of higher dimension less so. 
\subsubsection{Multiplying Constants}
The probabilities relating to the multivariate Wallenius distribution with multiple tables tend to be extremely small, that is of the order of $10^{-100}$ to $10^{-300}$, as noted earlier in Section \ref{s321}.  However, evaluating the logarithms of such  probabilities within the requisite calculations in R proved somewhat unstable. Therefore probabilities were  multiplied  by a constant which is approximately the inverse of the order of the probabilities, a strategy which has no impact on the calculations embedded in the likelihood and Bayesian settings.  Specifically, a multiplying constant does not affect the maximisation of the likelihood. In addition, the construction of pure likelihood intervals,  intervals based on Wilks' likelihood ratio statistic and the implementation of an MCMC algorithm  depend on  ratios of likelihoods and remain unaffected. This approach is used in the present study and logarithms of the likelihoods and of the associated probabilities are not introduced.  It should be emphasised here that  the use of likelihoods rather than  log-likelihoods is operational and is not in any way meaningful.
\subsection{A Markov Chain Monte Carlo Algorithm}\label{s42}
A Markov chain Monte Carlo (MCMC) algorithm was developed in the present study in order to accommodate the modelling of the multivariate Wallenius distribution within the Bayesian framework.  The algorithm is used in the examples which follow in the next two sections and is  presented generically here. Specifically, consider the posterior distribution of the unknown weight parameters $\bw$ expressed as  $[\bw | \bx] \propto \pi(\bw) L(\bx | \bw)$, where $L(\bx|\bw)$ is the likelihood of  $\bw$ with $\bx \sim W(\bm,n,\bw)$, $\pi(\bw)$ represents the pdf of a Dirichlet prior and, for compactness, the right hand term in the proportionality is denoted by $g(\bw)$. In essence, the MCMC introduced here is a Metropolis algorithm with the proposal implemented as a modified ball walk in Cartesian coordinates \citep{vempala:05} and the Metropolis filter in barycentric coordinates. 
The algorithm, termed the sphere walk Metropolis (SWM), is presented succinctly in Table~\ref{tab:algo} and   was programmed in the present study in R. The radius of the hypersphere, $r$, can be tuned to provide acceptance rates of the order of 25\% to 30\%. 
\section{The Wallenius Distribution with Three Categories}
In order to fix ideas with respect to the  fitting of the multivariate Wallenius distribution to data, $3 \times 2$  tables were taken as semi-synthetic data from the  journal preference dataset of \citet{graz:19}. 
Specifically, data from the three categories of journal with the smallest number of zero responses, that is Methodology,  Applied Statistics and Computation in that order, were  extracted from the $174$ responses.  There were no individuals for which all three counts were zero. The row marginal totals were again assumed to be those of the complete dataset and are given by  $\bm = (45, 34, 9)$ but the  samples sizes $n=\sum_{i=1}^3 x_i$, while taken to be fixed for each individual  $3 \times 2$  table, ranged from 6 to 20.  The three  counts of preferred journals in a given $3 \times 2$ table with fixed margins are denoted $\bx=(x_1,x_2,x_3)$ but, since $\sum_{i=1}^3 x_i=n$, only two are realisations of random variables.  Here the counts $x_1$ and $x_2$   are taken to be realisations of the random variables $X_1$ and $X_2$, the number of Methodological and Applied Statistics journals preferred. Finally, $\bx$ is distributed as  $W(\bm,n, \bw)$, with the weights $\bw=(w_1, w_2, w_3)$ conformable with $\bx$, that is with the order of  the categories of journal.   In addition, the constraints on the counts $0 \le x_1, x_2 \le n$  and $0 \le x_3=n-x_1-x_2  \le \min(n,9)$ necessarily  hold. 
The two main reasons for introducing the three-category Wallenius distribution are that the weights and attendant surfaces can be represented in two dimensions as ternary plots and that the MCMC algorithm introduced in Section \ref{s42} can be validated by simulation, albeit only for the examples in this section.
\subsection{A Single {\boldmath{$3 \times 2$} Table}}
\subsubsection{A Likelihood Lens}
Consider first  modelling the data for a single individual based on the Wallenius distribution with  three categories. The $174$ vectors of counts taken from the  $3 \times 2$  tables of interest clearly reflect the inherent imbalance in the data. Data for two individuals, one broadly representative of the overall responses and one somewhat atypical, were therefore selected for analysis. The row numbers of the individuals in the complete dataset and the attendant counts, with the counts for individual $20$ deemed typical and those for individual $108$ atypical, are presented  in Table\;\ref{tab:32ind}. The MLEs of the preference weights were obtained by invoking the unconstrained optimisation procedure outlined in Section \ref{s411} and are also included in that table. 

\begin{figure}
\vspace{-25mm}
\centering
\includegraphics[width=\textwidth]{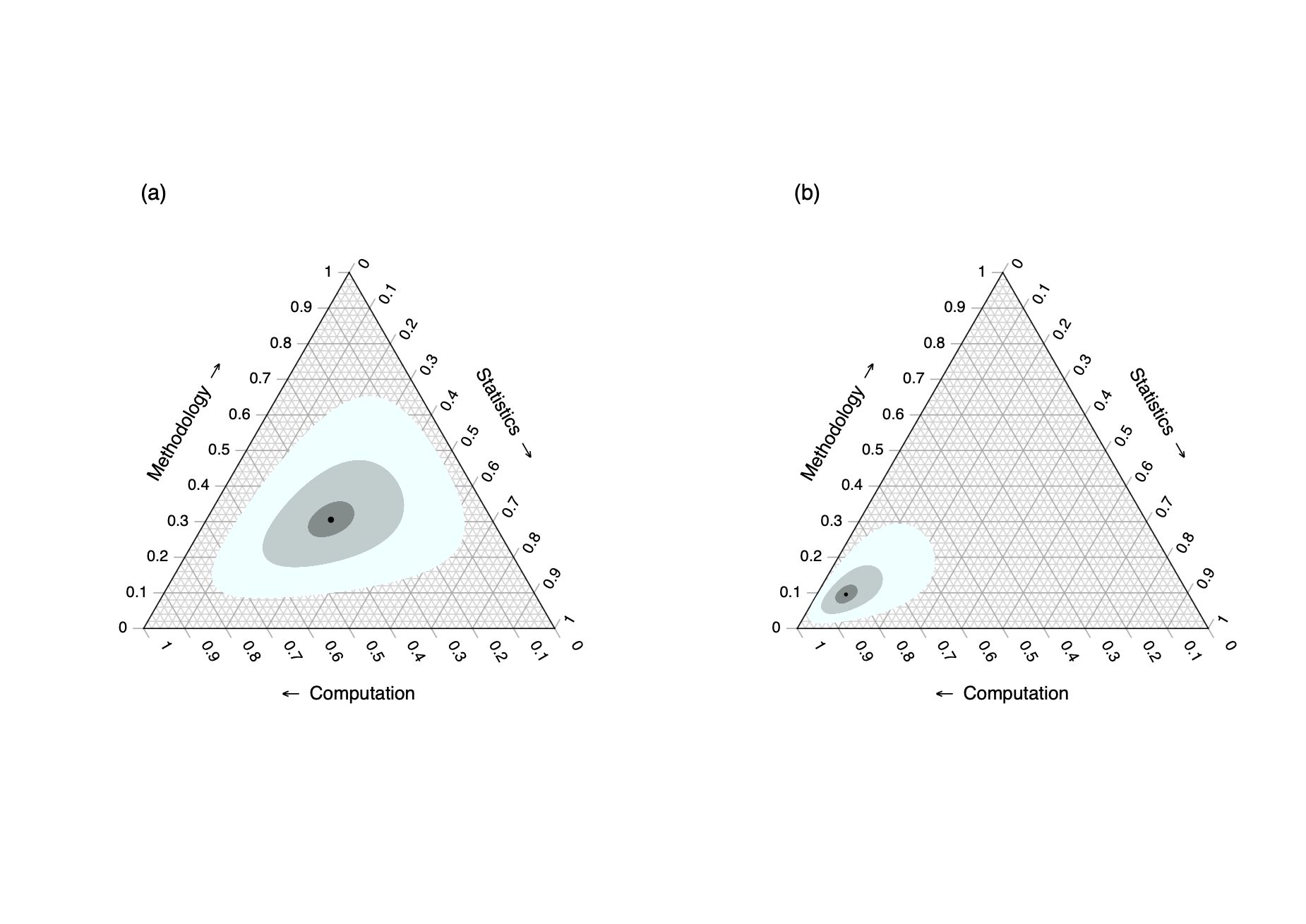} 
\vspace{-25mm}
\caption{ Plots of 95\%, 50\% and 5\% confidence regions for individuals 20 and 108}
\label{fig:wconf}
\end{figure}

The likelihood surfaces relating to the two $3 \times 2$  tables of interest, denoted $L(\bw | \bx)$,  were  simulated to a high degree of accuracy by taking a random sample of 1,000,000 uniform variables from the Dirichlet distribution, D$(1,1,1)$, over the simplex   and the  $95\%, 50\%$ and $5\%$  confidence regions based on Wilks' likelihood ratio statistic extracted from the simulations \citep{speagle:19}.  Plots of the likelihood surfaces so obtained, together with bivariate box plots delineating the specified confidence regions,  are displayed in Figure \ref{fig:wconf}. As an aside, the present  setting involves  two  unknown weight parameters and, as a consequence,  the confidence and pure likelihood regions coincide.

Confidence intervals for weights, such as those in the present study,  are not straightforward to construct and are, arguably, difficult to interpret  \citep{mostel:20}. As a consequence,  parametric bootstraps were conducted in order  to assess the accuracy of the MLEs of the individual weights. Specifically, both margins of the $3 \times 2$  tables  are  fixed and only $x_1$ and $x_2$, are realisations of random variables. The bootstrap replications thus comprise compositions of the integer $n$ into $3$ parts, with the constraint $x_3 \le 9$ then applied. For the two examples of interest here, the number  of  bootstrap samples is small and  ideal parametric bootstraps were conducted.  The numbers of bootstrap  samples, denoted  $\tilde{C}(n)$, and the values of  the bootstrap standard errors for the MLEs of the weights, denoted \seb, are presented in Table\;\ref{tab:32ind}. It is immediately clear from these latter values, and from the confidence regions displayed in Figure \ref{fig:wconf}, that the estimates of the attendant MLEs cannot be construed as being precise. 
\subsubsection{A Bayesian Lens}
Consider now a Bayesian framework for modelling data from a single $3 \times 2$  table based on the  Wallenius distribution with three categories. The posterior distribution of the weights, with a Dirichlet distribution as the prior, cannot be found explicitly. Recourse must therefore be made to simulating the posterior distribution over the simplex  in barycentric coordinates \citep{speagle:19} or to sampling the posterior distribution using an MCMC algorithm. 

To fix ideas, consider again individuals 20 and 108 from the journal preference data, with counts  $\bx$ of (8,6,2) and (7,4,7). Two priors, the flat prior,  D$(1,1,1)$, and an informative prior, D$(2,3,2)$, which reflects a sense that the Applied Statistics category should be more highly rated which and which aligns well with the likelihoods, were selected for  analysis. 
The posterior distributions were first approximated by  generating $1,000,000$ uniform random variates over the 2-dimensional simplex, as for the simulation of the likelihood,  and consistent results were obtained. 
The means and  standard deviations of the posterior distributions of the weights for the two priors are given in Table \ref{tab:32bayes} and the  MLEs of the weights for the two individuals are included in the legend of the table.  It is clear from a comparison of the posterior means with the corresponding MLEs  that the priors have a clear impact on the form of the likelihood.  

The posterior distribution of the weights was  also sampled using the SWM algorithm.   Three chains of length 100,000 were generated after a burn-in of 10,000 for each setting and mixing was found to be extremely fast. The trace plots and a multivariate potential scale reduction factor, $\hat{\mathrm R}$, of 1 indicated convergence of the chains.   The means and standard deviations so generated were found to be in excellent agreement with those obtained using simulation. The 95\% and 68\% credible intervals for the weights were also obtained and proved to be extremely wide in all cases. 
On balance therefore, the results within the Bayesian framework, together with those for the likelihood approach, underscore the fact that using the Wallenius distribution to model data for a single $3 \times 2$  table can be fragile and that the analysis of such data should be treated with caution. 
\subsection{Multiple {\boldmath{$3 \times 2$} Tables}}
Consider now the 174 responses to the categories Methodology, Applied Statistics and Computation, in that order.  Interest here focusses on fitting the multivariate Wallenius distribution  with a weight parameter, $\bw$, common to all individuals. The analysis is straightforward and  follows closely that of the single $3 \times 2$  tables.
\subsubsection{A Likelihood Lens} 
\begin{figure}
\vspace{-25mm}
\centering
\includegraphics[width=\textwidth]{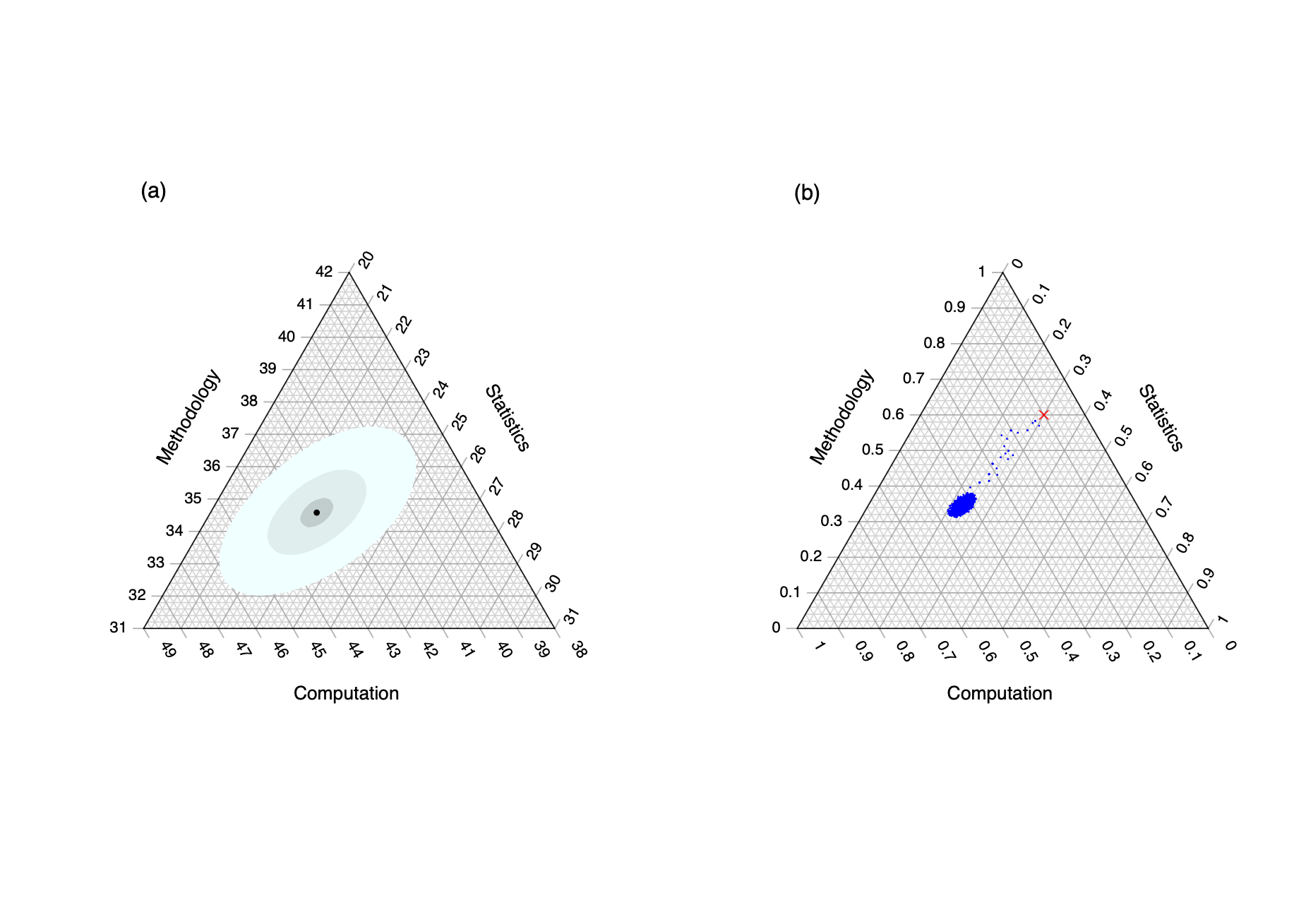}
\vspace{-25mm}

\caption{Multiple $3 \times 2$  tables. (a)  Likelihood regions: Magnified plots of 95\%, 50\% and 5\% confidence regions with $\wh=(0.346,0.228,0.426)$ (b) Bayes: Initial weights generated by the SWM algorithm with starting point $w_0=(0.6,0.3,0.1)$ a red cross.}
\label{fig:w32}
\end{figure}

The likelihood is given by $L(\bw|\bx_1, \ldots, \bx_{174}, \bn)=\prod_{i=1}^{174} L(\bw|\bx_i,n_i)$, where $\bx_i=(x_{i1}, \ldots, x_{i174})$, with values  of the order of $10^{-300}$ over a range of weights. As a consequence, the individual likelihoods embedded in the product term were multiplied by a factor of $50$ to provide stability in further calculations.  The  likelihood was found to be unimodal from a contour plot and the MLE of the weights was obtained as $\bwh= (0.346,0.228,0.426)$.  The 95\%, 50\% and  5\% confidence regions of the weights, magnified over a subset over a subset of the simplex, are shown as  bivariate box plots in Figure \ref{fig:w32}(a). A nonparametric  bootstrap comprising 5000 replications of  $\bwh$ was also conducted and yielded  the bootstrap standard errors \seb=(0.013, 0.011, 0.017) and the 95\% central percentile intervals   (0.321,0.371), (0.208,0.249), (0.391, 0.459) of the weights. It is immediately clear that mean values of the rating preferences with a weight common to all 174 individuals are far more precise than those of the single $3 \times 2$ tables.  
\subsubsection{A Bayesian Lens} 
The flat Dirichlet  distribution D$(1,1,1)$ was adopted here as a prior and the SWM algorithm of Section \ref{s42}  was again used to sample from the posterior distribution. The factor $\hat{\mathrm R}$ over three chains of 100,000 samples each was 1 and the trace plots indicated good convergence.  A plot of 10,000 samples taken from the start of one the the Markov chains at the point (0.6,0.3,0.1) is shown in Figure \ref{fig:w32}(b) and illustrates the fast convergence of the chain. The means, standard deviations, and  95\% credible intervals of the weights were found to be very close to the likelihood-based means, \seb~and 95\% percentiles and the credible regions were found to be very similar to those of confidence regions. As a consequence, details of the analysis within the Bayesian context are not recorded here.  
The probabilities that Computation was preferred to Methodology and Methodology to Applied Statistics were elicited from the SWM samples and are very close to one and exactly one, respectively,  thus indicating a clear order of preference for the three categories of journal.

The choice of a subjective prior for the present setting is not straightforward. More specifically, a good alignment of a Dirichlet prior distribution with the likelihood is not easy to find. For example, the prior D$(2,3,2)$, indicating that  Applied Statistics is underrated, is too diffuse and has little impact on the posterior distribution, while the more focussed prior D$(4,6,4)$, with a mean of $(0.286,0.428, 0.286)$ is totally misaligned with respect to the likelihood. 
The issue of finding a suitable prior to accommodate a sense that the estimated preferences do not fully reflect an opinion clearly requires considerable thought and is not pursued further  here. 
\section{Real World Examples}
\subsection{Categories of Journals} 
An analysis of the complete dataset of responses for the five categories of journals, Methodology,  Probability, Applied Statistics, Computation and Economics, based on the multivariate Wallenius distribution  is now  reported. Following  \citet{graz:19},  the vector of preference weights is taken to be common to all individuals.
The likelihood of the preference weights $\bw$ is given by $L(\bw|\bx,\bn)=\prod_{i=1}^{174} L(\bw|x_i,n_i)$ and has  values over a range of weights  of the order of $10^{-300}$. As a consequence, the individual likelihoods embedded in the product term were multiplied by a factor of $750$ which yielded  likelihood values of the order of $10^4$ and provided numerical stability in further calculations.   The MLE of $\bw$, denoted $\bwh$, was then obtained and a nonparametric  bootstrap of 2000 replications of  $\bwh$ was conducted, yielding bootstrap standard errors \seb~and the 0.25 and 0.975 percentiles points The results are summarised in Table\;\ref{tab:lik5}.

\begin{figure}[htbp]
\centering
\vspace{-25mm}
\includegraphics[width=\textwidth]{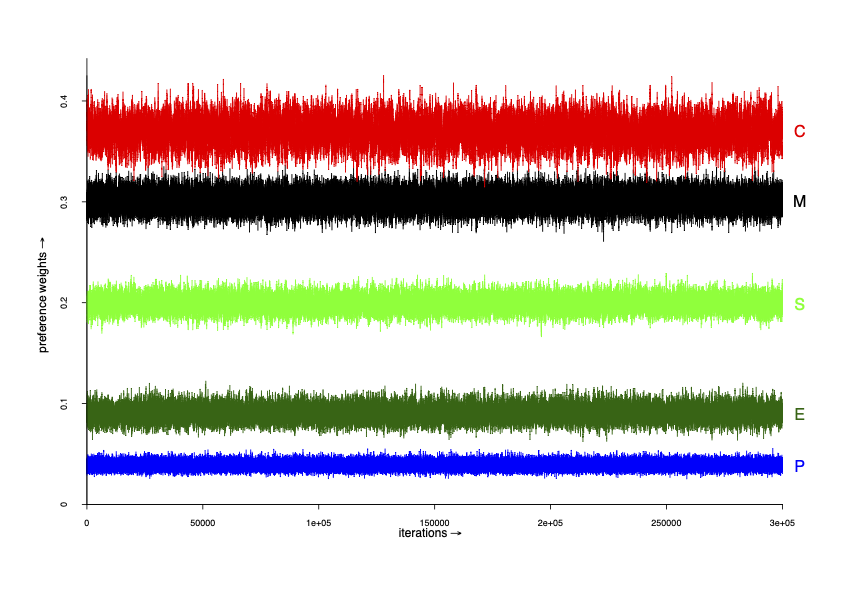} 
\vspace{-17.5mm}

\caption{Trace plots: 300,000 iterations of the SWM after burnout for the complete categories of journal dataset. C=Computation, M=Methodology, S=Applied Statistics, E=Economics and P=Probability}
\label{fig:onebig5}
\end{figure}

The complete  dataset for the five categories of  journals was also analysed within the Bayesian framework.  An objective Bayes setting with  the flat prior D$(1,1,1,1,1)$ was adopted and the posterior distribution was sampled using the SWM algorithm. Three chains of length 100,000 after a burn-in of 10,000, were so generated. The factor $\hat{\mathrm R}$ was equal to 1 and the trace plots, which are shown in Figure \ref{fig:onebig5}, indicated good convergence. The means and standard deviations coincided to within $\pm0.001$ with those of  the likelihood-based means and \seb~reported  in Table\;\ref{tab:lik5} and are not presented here.

On balance, the trace plots associated with the five categories of journal shown in  Figure \ref{fig:onebig5} illustrate, in essence, the key features of the objective Bayes analysis.  
The categories are clearly ordered  by decreasing preference as Computation, Methodology, Applied Statistics, Economics and Probability. Furthermore, the probability of preferring a category to that immediately lower in the ordering is exactly one, except for that of preferring  Computation to Methodology which is  0.99977.  The posterior distributions of the five weights, together with the mean and the 95\% and 68\% credible intervals, highlight the intrinsic variability between the categories but are not shown here for compactness. The choice of a subjective prior for the present setting is not straightforward and was not pursued further.

\citet{graz:19} present an analysis of  the dataset for the categories of journal based on objective Bayes. The authors deemed an MCMC approach to sampling the posterior distribution of the weights to be infeasible because of the intractability of the likelihood function and instead used approximate Bayesian computation (ABC). The results of \citet{graz:19} are broadly the same as those presented here in terms of the order of the categories of journal according to preference. However the means do not coincide exactly with those of the present study and the distributions of the weights are far more diffuse.
In addition, it is intriguing  to note that the means obtained from the paired comparisons of the categories and reported in Section \ref{s321} are close to those given here.
\subsection{Selective Predation}  
\citet{manly:74} reported an experiment  conducted  in order to investigate whether or not  eight types of fly survive food competition was dependent on the number of  their bristles. The data are presented in Table 4 of his paper. \citet[p.\,288]{manly:74} observed that, since the dead flies were not replaced, the data could be framed as an $8 \times 2$  table with fixed margins and, being unaware of Wallenius's PhD work, developed a model  for  `measures of selectivity' associated with each type of fly, together with a simple formula for calculating the measures. Later, \citet{chess:78} recognised that the data for the  flies can be modelled using a multivariate Wallenius distribution and noted that the MLEs of the weights agreed well with the estimated measures of Manly's findings. 

\begin{figure}[htbp]
\centering
\vspace{-0mm}

\includegraphics[width=\textwidth]{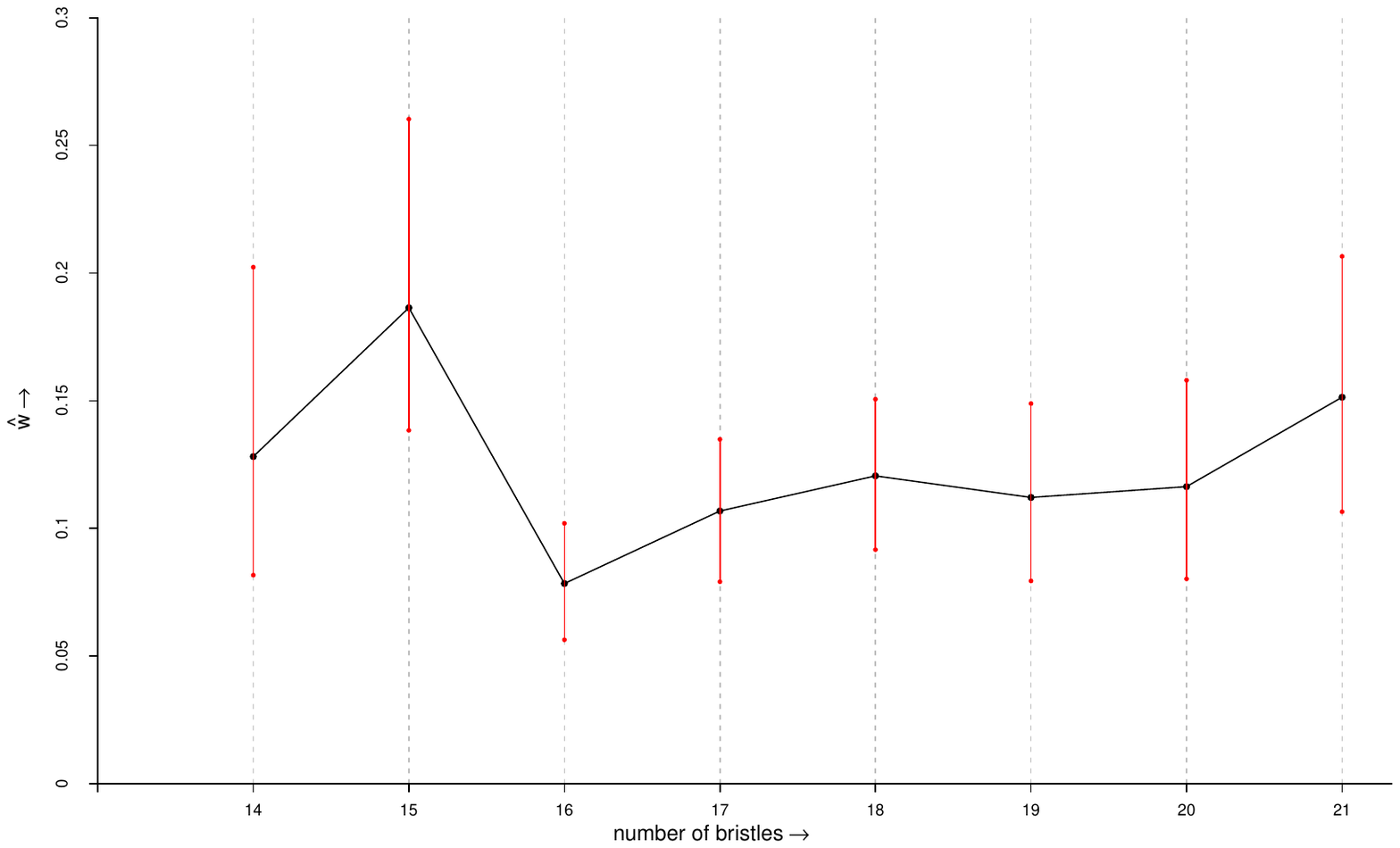} 
\vspace{-17.5mm}
\caption{Error bar plot based on the means and 95\% credible intervals obtained from the Bayesian analysis of the bristle fly data. }
\label{fig:bugs}
\end{figure}

The analysis of the data based on the Wallenius distribution was conducted here within both the likelihood and the objective Bayes frameworks. The vector of eight weights $\bw=(w_1, \ldots, w_8)$,  with $0 < w_i <1, i=1,\ldots,8$ and $\sum_{i=1}^8 w_i=1$, was associated with the numbers of  flies dying, in accord with the study of \citet{manly:74}. The MLE of the weights and the attendant  bootstrap standard errors and the 0.025 and 0.975 percentile points obtained from 2000 bootstrap samples, were computed.  In addition, a flat Dirichlet prior for the weights was introduced and the resultant posterior distribution sampled using  the SWM algorithm of the present study. The results of the likelihood-based and the Bayesian analysis  were similar and, for compactness, are not reported here.  Rather, the mean and the 95\% credible intervals of the weights obtained from the Bayesian analysis are displayed in an error bar plot in Figure \ref{fig:bugs}.
It is particularly interesting to observe that the means associated with the weights were very close to  those reported as normalised measures of selectivity by \citet{manly:74} and that the only new information presented here is the 95\% confidence intervals for the likelihood approach and the results for objective Bayes inference.
\section{Conclusions}
The results, and more particularly the examples, in this paper  are introduced  in order to demonstrate the ease with which  the Wallenius noncentral hypergeometric distribution can be fitted  to data emanating from a scheme in which sampling is sequential, biased and without replacement. The scheme is best represented  by a $c \times 2$ table  with the $c \ge 2$ rows corresponding to the categories of interest and the two columns relating to the numbers sampled and not sampled.
Likelihood-based and Bayesian analyses are presented and rely on the numerical calculation of  likelihoods  rather than  log-likelihoods. 
For $2 \times 2$ tables,  a preference weight,  and not the odds,  associated with the two categories sampled is adopted. The single parameter permits the approximation of integrals within the Bayesian context by simulation and  is easy to represent and interpret. In addition, inference for the weight  can  immediately be translated into that for an odds  or an odds ratio. 
For $c \times 2$  tables with more than two categories, that is $c>2$, some further numerical subtleties are  required. The unknown parameter is now taken to be a vector of preference weights which are positive and sum to one. Problems with respect to optimisation of  functions, such as the likelihood, with respect to the weights can be overcome by introducing a suitable transformation to yield an unconstrained optimisation problem. In addition, the likelihoods are generally small but can be scaled to obviate computational  underflow without changing the elements of inference. The MCMC algorithm, termed  the sphere  walk Metropolis algorithm, which is introduced for sampling from the requisite posterior distributions within the Bayesian framework, is novel and will be presented within a broader context in a separate paper.

The Wallenius distribution has undoubtedly been neglected in the statistical literature.  It can be argued that the integral embedded in the probability mass function of the distribution is algebraically intractable and  that the weight parameters lie on the regular simplex and are difficult to handle.  However  preference weights  are often  an  integral part of  models for  datasets  collected in areas ranging from ecology and finance to decision making.  It is hoped therefore that some of the problems associated with the requisite calculations for the Wallenius distribution  have been resolved in the present paper and that the nature of the examples presented here will lead to further studies.  \\

\noindent{\bf Conflict of interest} 

The author declares no competing interests.\\

\noindent{\bf Funding} 

I would like to thank the University of Cape Town and the National Research Foundation (NRF) of South Africa, Grant Number RA240111202320, for financial support. Any opinion, finding and conclusion or recommendation expressed in this material is that of the author and the NRF does not accept liability in this regard.\\ 

\noindent{\bf Acknowledgments} 

I would like to thank my two students, Usanda Mtwazi and Okuhle Nyangintsimbi,  for  having the courage to choose my project ``The Wallenius Distribution'' in their honours year at the University of Cape Town.  Their contributions during the early stages of this study were invaluable. I would also like to thank Birgit Erni and Murray Christian of my own department for many helpful discussions and Martin Smith of Durham University in the U.K. for help with his excellent R package {\tt Ternery}.

\bibliographystyle{chicago}   
\bibliography{wallenius}

\begin{table}[ht]
\caption{A $2 \times 2$  table with  $m_1, m_2$ and $n$ fixed. The table is completed by specifying the count of  red balls drawn, $x$.}
\label{tab:uni1}
  \centering
$$
\begin{array}{c|cc|c}
 &  \mathrm{Sampled} &  \mathrm{Not~Sampled} &  \mathrm{Total}  \\ \hline
 \mathrm{Red} & x & m_1 -x & m_1 \\
 \mathrm{White} & n-x & m_2 -(n-x) & m_2 \\ \hline
  \mathrm{Total}  & n & m_1+m_2 -n & m_1+m_2
 \end{array}
$$
\end{table}

 \begin{table}[ht]
\caption{A $c \times 2$ table with  $m_i, i=1, \ldots, c$ and $n$, fixed. The table is completed by specifying the counts $x_i$ of the the groups $i, i=1,\ldots, c-1$, and $x_c=n-\sum_{i=1}^{c-1}x_i$.}
\label{tab:multi}
  \centering
$$
\begin{array}{c|cc|c}
 \mathrm{Category}  &  \mathrm{Sampled} &   \mathrm{Not~Sampled} &  \mathrm{Total}  \\ \hline
1 & x_1 &  m_1-x_1 & m_1 \\
2 & x_2 & m_2 -x_2 & m_2 \\ 
\vdots & \vdots & \vdots  &\vdots  \\
c & x_c &  m_c-x_c & m_c \\ \hline
\mathrm{Total}  & n & M -n & M \\
 \end{array}
$$
\end{table}

 \begin{table}[h!]
  \centering
  \caption{(a) The $5 \times 2$ table of responses of the 20th individual taken from the complete dataset of Grazian, Leisen and Liseo (2019) and (b) the $2 \times 2$ table of the responses for Methodology and Applied Statistics extracted as semi-synthetic data from that table. The headings Yes  and No indicate Sampled and Not Sampled}
   \label{tab:gdata}  
$$
\begin{array}{l|cc|c}
 \mathrm{Category}  &  \mathrm{Yes} &   \mathrm{No} &  \mathrm{Total}  \\ \hline
\mathrm{Methodology} & 8 &  37 & 45 \\
\mathrm{Probability} & 0 & 23 & 23 \\ 
\mathrm{Statistics} & 6 & 28  & 34  \\
\mathrm{Computation} & 2 &  7 & 9 \\ 
\mathrm{Economics} & 1 &  12 & 13 \\ \hline
\mathrm{Total}  & 17 & 107 & 124 \\
 \end{array}
\hspace{20mm} \begin{array}{l|cc|c}
 \mathrm{Category}  &  \mathrm{Yes} &   \mathrm{No} &  \mathrm{Total}  \\ \hline
\mathrm{Methodology} & 8 &  37 & 45 \\
\mathrm{Statistics} & 6 & 28  & 34  \\  \hline
\mathrm{Total}  & 14 & 65 & 79 \\
 \end{array}
 $$
 \end{table}

\begin{table}[h!]
  \centering
  \caption{The dataset for the genetic study on the survival of neonate rabbits with phenotypes homozygote and heterozygote}
$$
\begin{array}{c|cc|c}
 &  \mathrm{Survived} &  \mathrm{Died} &  \mathrm{Total}  \\ \hline
 \mathrm{Homozygotes} & 41 & 54 & 95 \\
 \mathrm{~Heterozygotes} & 34 & 86 & 120 \\ \hline
  \mathrm{Total}  & 75 & 140 & 215
   \end{array}
$$
\label{tab:rabbits}
 \end{table}
 
\begin{table}[h!]
\caption{The mean, standard deviation, median and 95\% and 68\% credible intervals for the posterior distribution of the weight parameter $w$ for the  Wallenius data using  Bayesian inference with priors Be(1,1), Be(1,2) and Be(2,4).}
\label{tab:unibayes}
  \centering
    {  \addtolength{\tabcolsep}{-1mm}
      \begin{tabular}{|l|c|c|c|c|c|}
  \hline 
 & & & & \multicolumn{2}{|c|}{Credible Intervals} \\
Beta Prior &  Mean &   SD  &  Median  & 95\% & 68\% \\  \hline
Be(1,1) & $0.625$ & $0.054$  & $0.626$ & $(0.515, 0.726)$ & $(0.571,0.678)$\\ 
Be(1,2) & $0.617$ & $0.054$  &  $0.618$  & $(0.508, 0.718)$ & $(0.563, 0.671)$ \\
Be(2,4) & $0.607$ & $0.053$  & $0.608$  & $(0.499, 0.708)$ & $(0.553, 0.660)$  \\ \hline
\end{tabular}}
\end{table}

\begin{center}
\begin{table}[h!]
\caption{Paired Comparisons of categories of journals for the Grazian, Leisen and Liseo (2019) dataset.  The terms $T, n_{10}$ and $n_{01}$ represent the number of individuals compared and the number of  individuals preferring only Category 1 and  only Category 2, respectively, within the relevant comparison. The MLEs, pure likelihood intervals, \seb~and 95\% central percentile intervals for the weight appropriate $w$ are then given.}
\label{tab:pairs}
\vspace{5mm}
  \centering
    {  \addtolength{\tabcolsep}{-1mm}
      \begin{tabular}{|l|l|c|c|c|cc|c|c|c|}
  \hline 
& & & & & & & 95\% Likelihood & & 95\% Bootstrap
\\
Category $1$ &  Category $2$ & $T$ & $n_{10}$ & $n_{01}$  & $\wh$ & $1-\wh $ &  Intervals &\seb &  Intervals\\  \hline
Methodology  & Probability & 174 & 127 & 0 &  0.886 & 0.114 & (0.866,0.905)  & 0.017 & (0.852,0.919)\\  \hline
    Methodology  & Applied Stats & 174 & 8 & 0 & $0.602$ & $0.398$ &  (0.581,0.623) & 0.013 & (0.577,0.628)\\  \hline
 Methodology  & Computation & 174 & 31 & 0 & 0.448 & 0.552 & (0.420,0.477) & 0.018 & (0.414,0.485) \\  \hline 
Methodology  & Economics  & 174 & 93 & 0 & $0.770$ & $0.230$  &  (0.739,0.800)  &  0.020& (0.729,0.810) \\ \hline
Probability  & Applied Stats & 170 & 4 & 123 & $0.162$ & $0.838 $ & (0.136,0.191) & 0.024 & (0.116,0.212)\\   \hline
Probability  & Computation & 156 &13 & 109 & $0.094$ & $0.906 $ & (0.075,0.113) & 0.017 & (0.064,0.129)\\  \hline
Probability & Economics & 107 & 26 & 60 & 0.301 & $0.699$ &  (0.251,0.356) & 0.044& (0.217,0.390)\\ \hline
Applied Stats & Computation & 170 & 27 & 4 & $0.348 $ & $0.652 $ & (0.320,0.377) & 0.018 & (0.314,0.386) \\  \hline
Applied Stats & Economics & 168 & 87 & 2 & $0.689$ & $0.311$  & (0.650,$0.727)$  & 0.026 & (0.638,0.740) \\ \hline
Computation  & Economics &154  & 73 & 11 & $0.808$ & $0.192$  & (0.776,0.837) & 0.020 & (0.766,0.846) \\ \hline
 \end{tabular}}
\end{table}   
\end{center}

 \begin{table}[h!]
\caption{The Sphere Walk Metropolis Algorithm with Radius $r$ Fixed.}
\label{tab:algo}
\vspace{-2.5mm}

\begin{enumerate}
\item[1:] Choose a starting weight expressed in barycentric coordinates, $\bw_0$. Set $t=0$. 
\item[2:] Express the weight $\bw_t$ in Cartesian coordinates as $\bc_t$.
\item[3:]  Generate a uniform random point $\widetilde{\bc}$ on the hypersphere, that is on the boundary of the ball,  with centre  $\bc_t$ and fixed radius $r$. If the point $\widetilde{\bc}$ is not a weight, set  $\bw_{t+1} =\bw_t$ and go to Step 2; otherwise express the weight $\widetilde{\bc}$ in barycentric coordinates as $\widetilde{\bw}$. 
\item[4:] Accept the weight $\widetilde{\bw}$ w.p. min$\ds \bigg\{1, \frac{g(\widetilde{\bw})}{g(\bw_t)}\bigg\}$ and set $\bw_{t+1} =\widetilde{\bw}$; otherwise set $\bw_{t+1}= \bw_{t}$.
\item[5:] Increment $t$ and go to Step 2.
\end{enumerate}
\end{table}

\begin{table} [h!]
    \centering
\caption{The row numbers, the observed counts $\bx$, the MLEs of the weights $\bwh$, the numbers of samples in the ideal bootstrap $\tilde{C}(n)$ and the bootstrap standard errors \seb~obtained  for the two individual $3 \times 2$ tables of interest.}
\label{tab:32ind}
$$
\begin{array}{|c|c|c|c|c|c}
\hline
 \mathrm{Individual}  &  \bx &   \bwh &  \tilde{C}(n)  & \widehat{\mathrm {se}}_{\mathrm {boot}} \\ \hline
\mathrm{20} & ( 8, 6, 2) & (0.305, 0.303, 0.392) & 125  & (0.124, 0.127, 0.181)\\
\mathrm{108}  &(7, 4, 7) & (0.095,0.071,0.834) & 145 &  (0.055, 0.047, 0.085)\\   \hline
 \end{array}
$$
\end{table}

\begin{table} [h!]
    \centering
    \caption{Means and standard deviations for the posterior distributions of the weights with the priors D(1,1,1) and D(2,3,2)  for individuals 20 and 108. Note that the corresponding MLEs of the weights are $(0.305,0.303, 0.392)$ and $(0.096, 0.071, 0.834)$, respectively.} 
    \label{tab:32bayes}
 \begin{tabular}{|l|c|c|c|c|}
\hline 
 & \multicolumn{2}{|c|}{Individual 20}  &  \multicolumn{2}{|c|}{Individual 108} \\ \hline
Prior &  Mean &  SD   &  Mean &  SD  \\  \hline
D$(1,1,1)$ & $(0.324,0.319,0.357)$ & $(0.106,0.108,0.142)$ & $(0.134, 0.107, 0.759)$ & $(0.060,0.055,0.091)$ \\  \hline
D$(2,3,2)$ & $(0.320,0.351, 0.329)$ & $(0.095, 0.101, 0.121 )$ & $(0.168, 0.162, 0.670)$ & $(0.065, 0.066, 0.101)$ 
 \\  \hline
\end{tabular}
\end{table}

\begin{table}[h!]
\caption{MLEs, \seb~\,and the 95\% central confidence intervals of the weights $w_1, \ldots, w_5$ in the order Methodology,  Probability, Applied Statistics, Computation and Economics  for the analysis of  the complete dataset of Grazian, Leisen and Liseo (2019)}
\label{tab:lik5}
  \centering
    {  \addtolength{\tabcolsep}{-1mm}
      \begin{tabular}{|l|c|c|c|c|c|}
  \hline 
 Metric &  $w_1$ &  $w_2$ & $w_3$ & $w_4$ & $w_5$ \\  \hline
MLE  & 0.302 & 0.038 & 0.199 & 0.371& 0.090\\  \hline
\seb & 0.010 & 0.006 & 0.009 & 0.016 & 0.009\\ \hline
95\% intervals & (0.282,0.323) & (0.027,0.051) & (0.182,0.218)    & (0.340,0.403)  & (0.073,0.109) \\ \hline
\end{tabular}}
\end{table}

\end{document}